\documentclass[graybox]{svmult}

\usepackage{mathptmx}       
\usepackage{helvet}         
\usepackage{courier}        
\usepackage{type1cm}        
\usepackage{makeidx}         
\usepackage{graphicx}        
\usepackage{multicol}        
\usepackage[bottom]{footmisc}

\makeindex             

\usepackage[show]{sssrced}
\usepackage{amssymb}
\newcommand{\approxsim}{\sim}

\begin{document}

\title*{Prospects for asteroseismology of rapidly rotating B-type stars}
\author{Hideyuki Saio}
\institute{Hideyuki Saio \at Astronomical Institute, Graduate School of Science, Tohoku University, Sendai, Japan \email{saio@astr.tohoku.ac.jp}
}
%
%
\def\bm#1{\mbox{\boldmath$#1$}}

\maketitle

\abstract{
In rapidly rotating stars Coriolis forces 
and centrifugal deformations modify 
the properties of oscillations; the Coriolis force 
is important for  low-frequency modes, while the centrifugal 
deformation affects mainly p-modes.
Here, we discuss properties of g- and r-mode oscillations in rotating stars.
Predicted frequency spectra of high-order g-modes (and r-modes) excited 
in rapidly 
rotating  stars show frequency groupings associated with azimuthal order $m$.
We compare such properties with observations in rapidly 
rotating Be stars, and discuss what is learnt 
from such comparisons.
}

\section{Oscillations in main-sequence B-stars}
\begin{figure}[t!]
\begin{center}
\includegraphics[scale=.4]{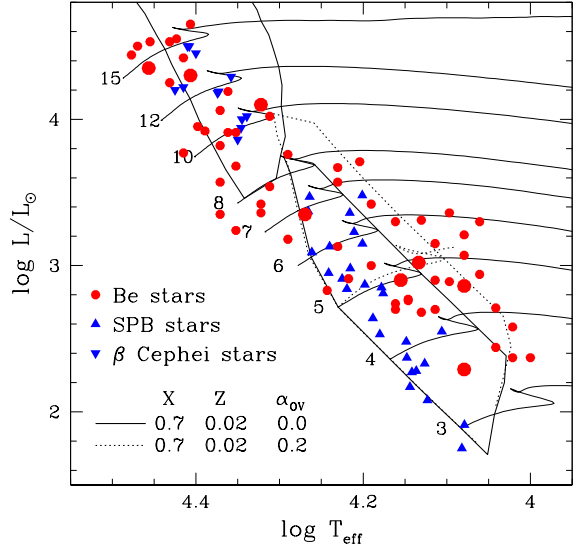} 
\end{center}
\caption{
Positions of rapidly rotating Be stars (filled circles) in the HR diagram are shown along 
with the well known B-type main-sequence variables SPB (triangles) and $\beta$ Cep stars 
(inverted triangles).
Big filled circles indicate the Be stars with nonradial pulsations detected by space 
photometry from  the MOST and CoRoT satellites. 
Parameters of Be stars are taken mainly from \cite{fre06,zor05},  and of SPB stars 
from \cite{dec02,nie02}.
Also shown are theoretical instability boundaries for p- and g-modes.
Solid lines for evolutionary tracks and instability boundaries are from models without
core overshooting, while dotted lines come from models with a core overshooting of $0.2H_p$
($H_p$=pressure scale height).
Theoretical models were obtained using a standard chemical composition of 
$(X,Z)=(0.70,0.02)$ with OPAL opacity tables \cite{igl96}.
}
\label{fig:hrd}     
\end{figure}

Thanks to  OPAL and OP opacity tables \cite{igl96,bad05}, 
we now understand that 
radial and nonradial oscillations found in main-sequence B-stars;\,i.e., $\beta$ Cephei and SPB
(Slowly pulsating B) stars,  are excited by the 
kappa-mechanism associated with the Fe opacity bump at $T\sim2\times10^5$K
\cite{kir92,mos92,gau93,dzi93}. 
Low-order p- and g-modes are excited in $\beta$ Cephei stars,
while high-order g-modes are excited in SPB stars.
Figure\,\ref{fig:hrd} shows their positions in the HR diagram and predicted instability 
regions; solid and dotted lines are for models without and with core overshooting,
respectively. 
The instability regions bounded by solid lines are 
roughly consistent with $\beta$ Cephei (inverted triangles) and SPB (triangles) stars, 
which are mostly slow rotators.

Stellar oscillations give us useful information on the stellar interior which is 
hard to obtain by other means.
For $\beta$ Cephei stars having low-order p- and g-modes, mode identifications
are less ambiguous so that detailed asteroseismic studies are possible. 
Comparing observed frequencies with theoretical ones yields best estimates of
physical parameters
as well as the extent of core-overshooting for each star.
In addition, rotational $m$-splittings of p- and g-modes, which have different depth 
sensitivity,  can be used to measure the
strength of differential rotation in the stellar interior.
Such asteroseismic analyses have been done for some $\beta$\,Cephei stars; e.g., 
$\nu$\,Eri (\cite{aus04,dzi08,sua09,das10}), $\theta$\,Oph (\cite{bri07,lov10}),
 HD\,129929 (\cite{dup04}),  $\beta$\,CMa (\cite{maz06}), 12\,Lac (\cite{des09}), 
 and $\delta$\,Cet (\cite{aer06}).
Although results are still somewhat controversial, they seem to indicate that
the extent of core overshooting in $\beta$\,Cephei stars ranges around 
$0.1$--$0.3H_p$,
and the core-to-envelope ratios of rotation rates are approximately 
$3$--$5$. 

In contrast to the cases of $\beta$ Cephei stars, detailed seismic analysis for SPB stars 
is difficult,  because g-mode frequencies are densely distributed, 
affected strongly by stellar evolution, and modified significantly even by moderate stellar
rotation; these properties make mode identifications difficult 
(see De Cat~\cite{dec07} for a review on SPB stars).

Although a detailed frequency fitting between theory and observation 
for each SPB star might be difficult, a collective property as simple as 
the distribution of SPB stars in the HR diagram
gives us useful information on the extent of convective-core overshooting. 
The ``red" (or cooler) boundary of the SPB (g-mode) instability region 
corresponds to the disappearance of convective cores.
The Brunt-V\"ais\"al\"a frequency in a radiative dense core is very high 
and hence the wavelengths of a g-mode become very short there,
which, in turn, cause strong radiative damping \cite{dzi71, gau93,dzi93}.
This is the reason why the cool boundary from models with a core overshooting of
$0.2H_p$ (dotted lines in Fig.\,\ref{fig:hrd}) is redder than that from models without 
core overshooting (solid lines).
Figure\,\ref{fig:hrd} shows that the observed SPB stars lay within the instability 
boundary obtained from models without core overshooting.
This indicates that, in contrast to the cases of $\beta$ Cephei stars, 
mixing by core overshooting should be weak in SPB stars; 
i.e., in the slowly rotating main-sequence stars of $3$--$8M_\odot$. 
We need further accumulation of fundamental parameter data of SPB stars to confirm
the property.

\begin{figure}[t!]
\hspace{-0.02\textwidth}
\includegraphics[scale=.3]{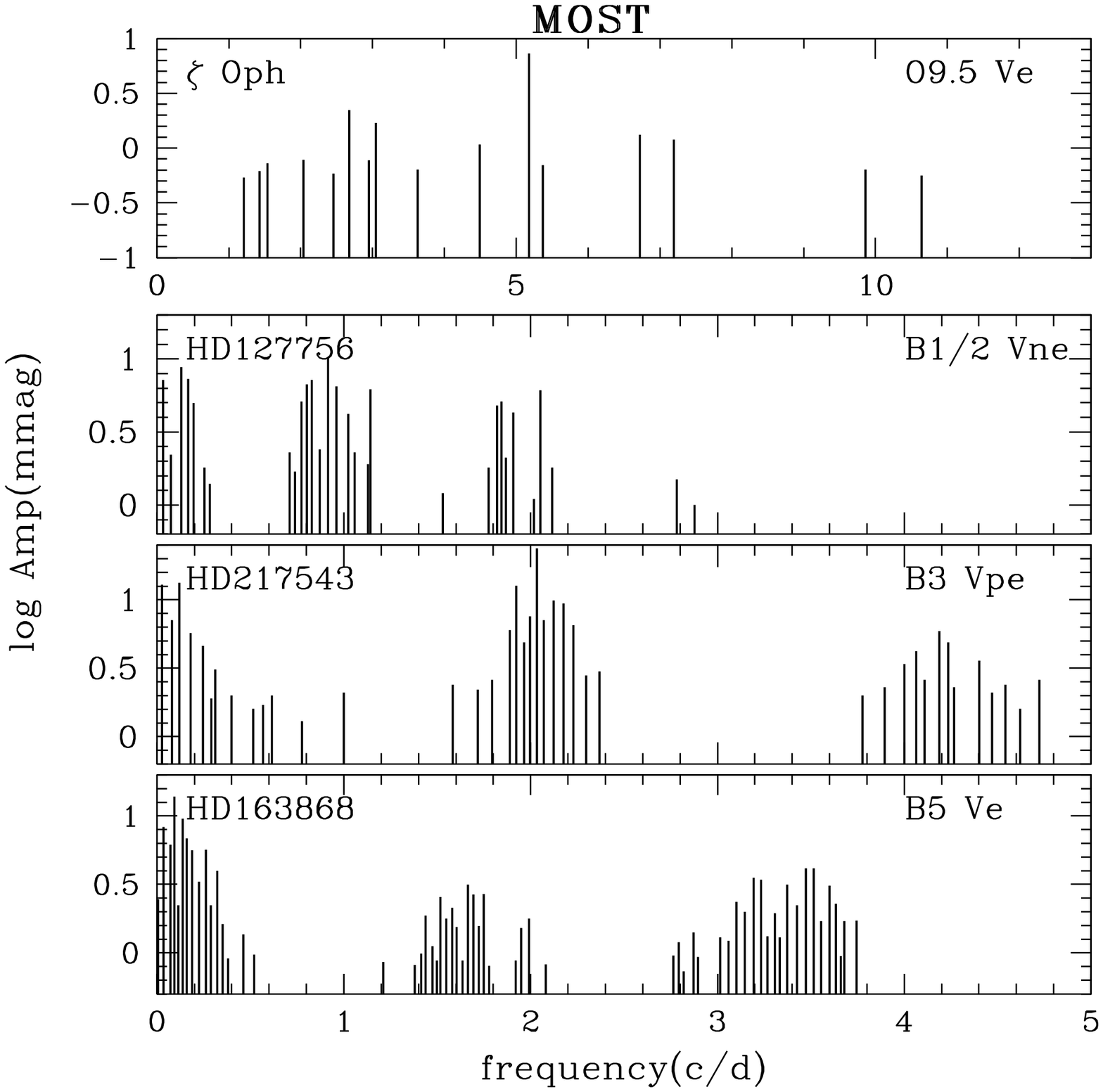}
\includegraphics[scale=.3]{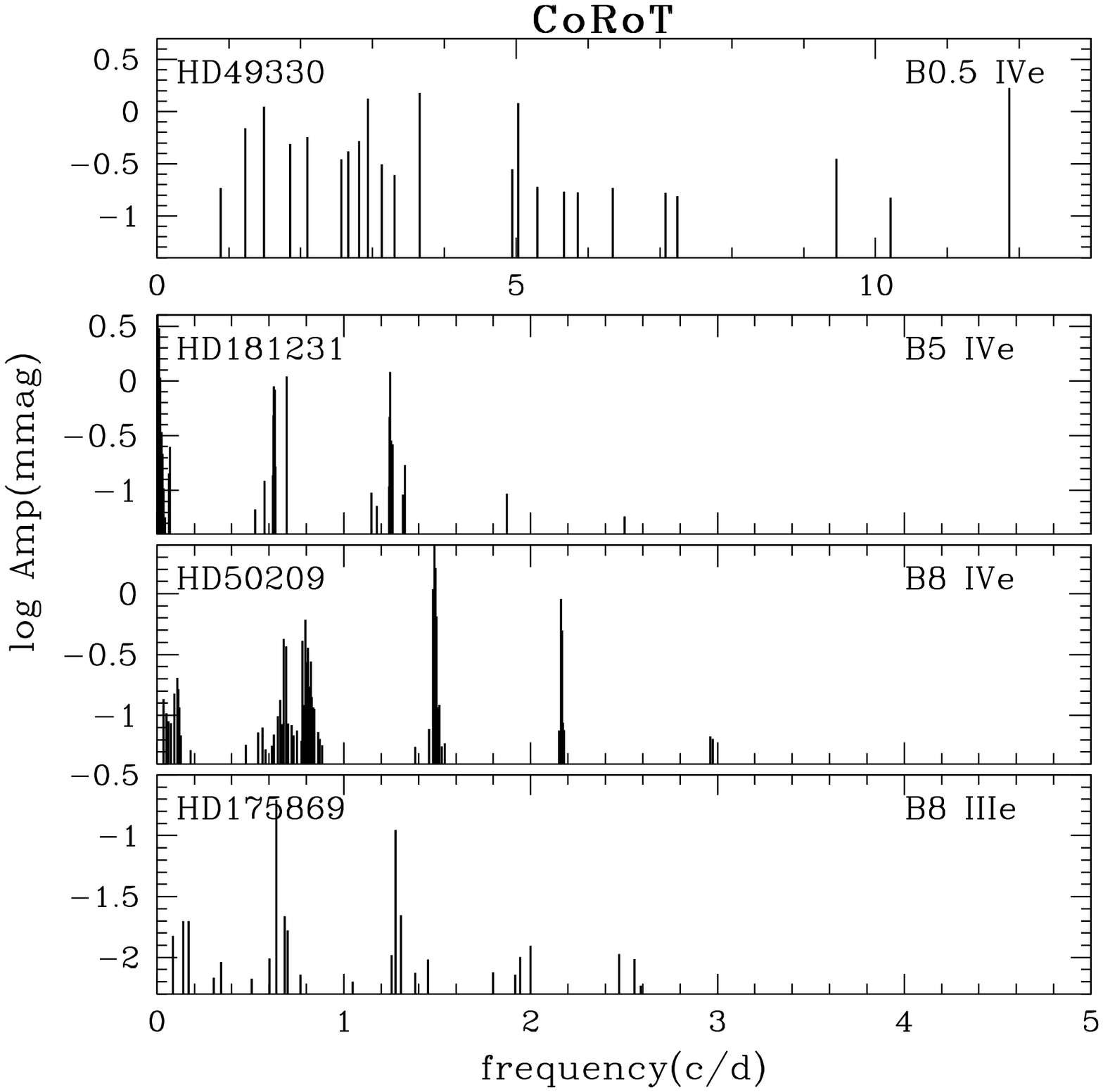}
\caption{
Frequency-amplitude diagrams for light variations detected in Be stars by
the MOST (left panel) and the CoRoT (right panel) satellites.
In addition to these stars, MOST detected in the late-type Be star $\beta$ CMi 
small amplitude oscillations with frequencies of  3.26 and 3.28 c/d \cite{sai07}.
Frequencies and amplitudes adopted from \cite{wal05a}--$\zeta$ Oph, 
\cite{cam08}--HD 127756 \& HD 217543, \cite{wal05b}--HD 163868,
\cite{hua09}--HD49330, \cite{nei09}--HD 181231,\cite{dia09}--HD 50209,
and \cite{gut09}--HD 175869. 
Note that the stars observed by CoRoT tend to be more evolved 
(with the spectral luminosity class IV) than the
stars observed by MOST.
}
\label{fig:befreq}
\end{figure}

Also plotted in Fig.\,\ref{fig:hrd} are the positions of rapidly rotating Be stars 
(filled circles).
Be stars are B-type stars which have (or had sometime before) Balmer lines in emission.
The emission lines arise from a circumstellar disk ejected from the star due to rapid rotation.
Many Be stars are known to show spectroscopic and photometric variations on 
various time-scales; in particular, short-term (order of a day) photometric 
(e.g., Balona~\cite{bal95}) and line-profile (e.g., Rivinius \& Baade~\cite{riv03}) 
variations are thought to be caused by radial 
and nonradial pulsations of Be stars, although rotational modulations were also suggested 
for the photometric variations.   The Be stars with short-term photometric variations are 
sometimes called $\lambda$ Eri variables (\cite{bal95}).
Interestingly, Be stars tend to be located in the $\beta$\,Cephei (p-mode) and 
SPB (g-mode) instability regions, supporting pulsation (oscillation) origin 
of the short-term variations.
Recently, the MOST and CoRoT satellites have found 
multi-periodic light variations in several Be stars, of which amplitude diagrams are
shown in Fig.\,\ref{fig:befreq}. 
The muliti-periodicity strongly supports the explanation 
that the short-term variations of Be stars
are caused by pulsations rather than rotational modulations.

The frequency spectra in the top panels of Fig.\,\ref{fig:befreq} 
for the hot Be stars, $\zeta$ Oph and HD 49330, show wide ranges of 
periodicities ranging from about a day to a few hours, while the lower panels show 
relatively long and grouped periods.
The difference can be understood by the fact that the hottest two stars lay in the 
$\beta$ Cephei instability range (Fig.\,\ref{fig:hrd}) and p-modes and low order g-modes
are excited, while the other relatively cooler Be stars lay in the SPB instability range 
where long-period g-modes are excited. 
Figure\,\ref{fig:hrd} shows that some cool Be stars including some whose variability are 
detected by CoRoT (large filled circles) lay outside of the SPB instability range obtained
from models without convective core overshooting. 
This indicates that an extensive mixing around the convective core should be occurring,
if nonradial g-modes are excited in those Be stars.
The analyses for HD 181231 and HD 175869 by Neiner et al.\cite{nei12} do indicate
the presence of such mixing.
Future observation of other cool Be stars should tell whether such
extensive mixing is ubiquitous in rapidly rotating Be stars.

In contrast to the broad frequency spectra of the hottest Be stars 
(top panels of Fig.\,\ref{fig:befreq}) without other remarkable features,
the spectra of the relatively cooler Be stars show conspicuous 
frequency groups which are regularly separated.
This property comes from strong rotation effects on high-order g-modes whose 
intrinsic frequencies are comparable with or less than the rotation frequencies.

\section{Oscillations in rotating stars}
Stellar rotation affects the oscillations in two ways; through the centrifugal force which deforms 
the equilibrium structure from spherical symmetry, and the Coriolis force which represents
the angular momentum conservation when matter moves.
In this section we discuss rotation effects mainly on low-frequency modes 
(for p-modes see e.g., Goupil~\cite{gou09}).

\subsection{Coriolis and centrifugal force effects} 
The effects of Coriolis and centrifugal forces can be seen from the equation of 
motion, which may be written in the inertial frame as
\begin{equation}
{\partial \vec{v} \over\partial t} + \vec{v}\cdot\nabla\vec{v} = -{1\over\rho}\nabla p - \nabla\psi ,
\label{eq:momentum}
\end{equation}
where $\vec{v}$ is fluid velocity, $\rho$ matter density, $p$ pressure, 
and $\psi$ gravitational potential. 
The velocity consists of oscillation velocity $\vec{v}'$ and rotation velocity
$r\sin\theta\Omega\vec{e}_\phi$, where $\vec{e}_\phi$ is the unit vector of  $\phi$ direction.
In equilibrium state without oscillations, 
Equation~(\ref{eq:momentum}) is reduced to
\begin{equation}
-r\sin\theta\Omega^2\vec{e}_\varpi = -{1\over\rho_0}\nabla p_0 - \nabla\psi_0,
\label{eq:equil}
\end{equation}
where 
\begin{equation}
\vec{e}_\varpi \equiv \sin\theta\vec{e}_r + \cos\theta\vec{e}_\theta
= \vec{e}_\phi\times\vec{e}_z , 
\end{equation}
is the outwardly pointing unit vector perpendicular to the rotation axis.
 
Writing variables in Equation\,(\ref{eq:momentum}) as a sum of the equilibrium value and 
(Eulerian) perturbation
due to oscillations; e.g., $p = p_0 + p'$, subtracting  Equation (\ref{eq:equil}) and 
disregarding non-linear terms with respect
to the perturbed quantities, we obtain
\begin{equation}
{\partial \vec{v}' \over\partial t} + \Omega{\partial\over\partial\phi}\vec{v}'
+\vec{v}'\cdot\nabla(r\sin\theta\Omega\vec{e}_\phi) 
= -{1\over\rho}\nabla p' +{\rho'\over\rho^2} \nabla p- \nabla\psi' ,
\end{equation}
where the subscript $_0$ for equilibrium quantities is dropped for 
ease of notation. 

The relation between oscillation velocity and the Lagrangian displacement $\vec{\xi}$ is
somewhat different in a rotating star.  
Since $\D\vec{\xi}/\D t$ corresponds to the Lagrangian velocity
of oscillation, the Eulerian oscillation velocity $\vec{v}'$ is written as
\begin{equation}
\vec{v}'= {\D\vec{\xi}\over \D t} - \vec{\xi}\cdot\nabla(r\sin\theta\Omega\vec{e}_\phi)
={\partial\vec{\xi}\over\partial t} + \Omega{\partial\vec{\xi}\over\partial\phi}
- \vec{\xi}\cdot\nabla(r\sin\theta\Omega\vec{e}_\phi).
\label{eq:vdash}
\end{equation} 
We express the temporal and azimuthal dependence of an oscillation mode as 
\begin{equation}
\E^{\I(\sigma t + m\phi)},  
\label{eq:temp}
\end{equation}
where $\sigma$ is the oscillation frequency observed in the inertial frame 
and $m$ is the  azimuthal order of the oscillation.  Equation (\ref{eq:temp})  
means that we adopt  the convention that prograde modes correspond to $m<0$; 
i.e., an oscillation propagates in the positive direction of $\phi$ (or in the 
direction of rotation velocity) if $m<0$. 
Then, Eq.~(\ref{eq:vdash}) becomes
\begin{equation}
\vec{v}'  =
\I\omega\vec{\xi} - \I\sigma r\sin\theta\vec{e}_\phi \vec{\xi}\cdot\nabla\Omega,
\label{eq:vdash2}
\end{equation}
where 
\begin{equation}
\omega\equiv\sigma+m\Omega 
\label{eq:omega}
\end{equation}
is the frequency in the frame rotating with an angular frequency of $\Omega$.
Note that in the case of uniform rotation, 
$\omega$ is constant in the stellar interior and can be adopted as eigenfrequency.  

Substituting this expression of $\vec{v}'$ into Equation~(\ref{eq:momentum}), we obtain
\begin{equation}
-\omega^2\vec{\xi}+2\I\omega\Omega(\vec{e}_z\times\vec{\xi})
+r\sin\theta(\vec{\xi}\cdot\nabla\Omega^2) \vec{e}_\varpi
= -{1\over\rho}\nabla p' +{\rho'\over\rho^2}\nabla p - \nabla\psi',
\label{eq:linmom}
\end{equation}
where 
$\vec{e}_z$ is the unit vector parallel to the rotation axis.
The second term in the left hand side, $2\I\omega\Omega(\vec{e}_z\times\vec{\xi})$, 
corresponds to the Coriolis force. 
The centrifugal-force effect is hidden in $\nabla p$, which is, in the non-rotating case, 
equal to $-\vec{e}_r GM_r\rho/r^2$.  In the presence of rotation the centrifugal force modifies 
$\rho^{-1}\nabla p$ by the order of $r\sin\theta\Omega^2 \vec{e}_\varpi$ 
and hence breaks spherical symmetry.
Thus, the importance of the centrifugal force is measured as 
$f_{\rm cen}\equiv r^3\sin\theta\Omega^2/GM_r$, 
which indicates that the effect of centrifugal forces is  largest at the equator on the 
stellar surface where $f_{\rm cen} = R_{\rm eq}^3\Omega^2/GM < 1$
with $R_{\rm eq}$ being the equatorial radius, 
and minimum near the center where 
$f_{\rm cen} \rightarrow {4\pi\over3}\sin\theta\Omega^2/G\rho_{\rm c} \ll 1$ with
$\rho_{\rm c}$ being the central density.

On the other hand, the importance of the Coriolis force can be measured by
\begin{equation}
f_{\rm cor} \equiv {2\Omega\over\omega},
\label{eq:coriolisparameter}
\end{equation}
which determines the relative importance between the first two terms on the left hand side
of equation (\ref{eq:linmom}); when $f_{\rm cor}>1$ the Coriolis term is larger than 
the acceleration term (first term) and {\it vice versa}.
Obviously, for a given rotation frequency g-modes are affected more strongly by Coriolis force  than p-modes because frequencies of p-modes in the co-rotating frame are larger than those
of g-modes.

The p-mode frequencies  are bounded as 
$\omega^2 \gtrsim 10GM/R^3$  (see e.g., Cox~\cite{cox80}), 
while the rotation frequency is limited by $\Omega^2 < GM/R^3$.
Therefore, for p-modes $f_{\rm cor} < 1$; i.e., the effects of Coriolis forces are always 
small for p-modes, while the effect of centrifugal force (deformation) can be significant
for p-modes because the amplitudes of p-modes are confined to outer layers. 
Actually, Reese et al.~\cite{ree09} 
have shown that p-mode properties are significantly
modified in rapidly rotating deformed stars.

For g-modes, on the other hand, the opposite is true; 
$f_{\rm cor}$ can be larger than unity even in a slowly rotating star for high-order g-modes,
while the effect of centrifugal force is small even in rapidly rotating stars because
the amplitude of g-modes is confined to inner layers where the centrifugal deformation is
small.
The insensitivity of g-mode frequencies to the centrifugal deformation has been shown
numerically by Ballot et al.~\cite{bal11} .
 
As seen in Equation (\ref{eq:linmom}), in the {\it absence} of rotation 
the horizontal displacement $\vec{\xi}_{\rm h}$ is proportional to 
$\nabla_{\rm h}(p'/\rho+\psi')$ with
\begin{equation}
\nabla_{\rm h}\equiv \vec{e}_\theta{\partial\over\partial\theta} 
+ \vec{e}_\phi{1\over\sin\theta}{\partial\over\partial\phi},
\end{equation}
which makes the angular dependence of an oscillation mode  
representable by a single spherical harmonic $Y_\ell^m(\theta,\phi)$  
(see, Cox~\cite{cox80}, Unno et al.\ \cite{unn89}, or Aerts et al.\ \cite{aer10} for details).
This simple property is lost in the presence of rotation even if the centrifugal deformations
are neglected, because of the presence of the Coriolis force.
Then, we express the angular dependence of oscillation 
by using a sum of terms associated with spherical harmonics as
\begin{equation}
\vec{\xi} = \sum_{j=1}^J\left[S^j Y_{l_j}^m\vec{e}_r + H^j \nabla_{\rm h}Y_{l_j}^m
+ T^j\left(\nabla_{\rm h}Y_{l'_j}^m\right)\times\vec{e}_r\right] \quad {\rm and} \quad
p'=\sum_{j=1}^Jp'^jY_{l_j}^m,
\label{eq:eigenfunc}
\end{equation}
where $l_j=\left|m\right|+2(j-1) + I$ and $l'_j=l_j+1-2I$  
with $I=0$ for even modes and $I=1$ for odd modes, 
and $J$ means a truncation length.
Other scalar variables are expressed in a way similar to $p'$.
The eigenfunction for a scalar variable of an even (odd) mode which consists of terms 
proportional to $Y_{l_j}^m$ with even (odd) values of ($l_j-\left|m\right|$) is 
symmetric (anti-symmetric) about the equatorial plane.

As seen in the above equations, we can still assign azimuthal degree $m$ and 
the (even or odd) parity of a mode, but no longer a latitudinal degree.
To identify the latitudinal behavior of a g-mode we sometimes use in this paper 
the notation $\ell_0$; the latitudinal degree the mode 
would have when $\Omega \rightarrow 0$.
When we discuss the observational properties, we use the effective 
degree $\ell$  which represents the value of $l_j$ for the component with the maximum 
amplitude on the stellar surface among the other components in Eq.\,(\ref{eq:eigenfunc}), 
and  we use $\ell'$ for the corresponding toroidal component.

Note that toroidal components are needed in representing the displacement vectors, 
because the toroidal velocity fields couple, through the Coriolis force, with spheroidal
velocity fields which generate variations in density, pressure, temperature, etc.

\subsection{r-modes}
Purely toroidal motions in a non-rotating spherical star do not disturb stellar structure 
so that no restoring force nor oscillations arise.
In the presence of rotation, however, the latitudinal gradient of the vertical component of 
the angular frequency of rotation $\Omega\cos\theta$ causes 
a restoring 
force for toroidal motions so that toroidal oscillations occur; 
we call these oscillations r-modes (or global Rossby modes). 
All the r-modes are retrograde in the co-rotating frame 
(see e.g., Pedlosky~\cite{ped79}, Saio~\cite{sai82}).
If we assume that the displacement is purely toroidal and neglect the centrifugal deformation,
the toroidal oscillation would have the 
limiting frequency, $\omega_{\rm rlim}$ given as
\begin{equation}
\omega_{\rm rlim} \equiv {2m\Omega\over \ell'(\ell'+1)},
\label{eq:rlim}
\end{equation}
in the co-rotating frame.
Actual frequencies of r-modes in the co-rotating frame deviate 
from $\omega_{\rm rlim}$; the deviation is larger for higher radial order modes
(Provost et al.\ \cite{pro81}, Saio~\cite{sai82}).
The deviation arises  because the Coriolis forces associated with
toroidal motion generate spheroidal motions; i.e.,
vertical motion and density perturbations are generated, and hence buoyancy 
plays a role in the restoring forces for r-modes.
Because of these effects, r-modes are sometimes called ``mixed modes" 
(Townsend~\cite{tow05})  or ``q-modes" (quasi g-modes; Savonije~\cite{sav05}).
The coupling with spheroidal motions makes it possible for r-modes (mainly 
toroidal oscillations) to be excited thermally by the kappa-mechanism in the 
same way as g-modes are excited.
Stability analyses by Townsend~\cite{tow05}, Savonije~\cite{sav05}, 
and Lee~\cite{lee06}, indicate that some r-modes are actually  
excited by the kappa-mechanism in intermediate mass (around $3$--$8M_\odot$) 
main sequence stars, in which g-modes are also excited (SPB stars).

\subsection{Latitudinal amplitude distributions}
Latitudinal distributions of oscillation amplitudes are expressed 
as shown in Equation\,(\ref{eq:eigenfunc}), 
which indicates a dependency on the ratio $f_{\rm cor}$. 
The distributions are important 
in understanding the visibilities of excited modes.
\begin{figure}[b!]
\hspace{-0.02\textwidth}
\includegraphics[scale=.3]{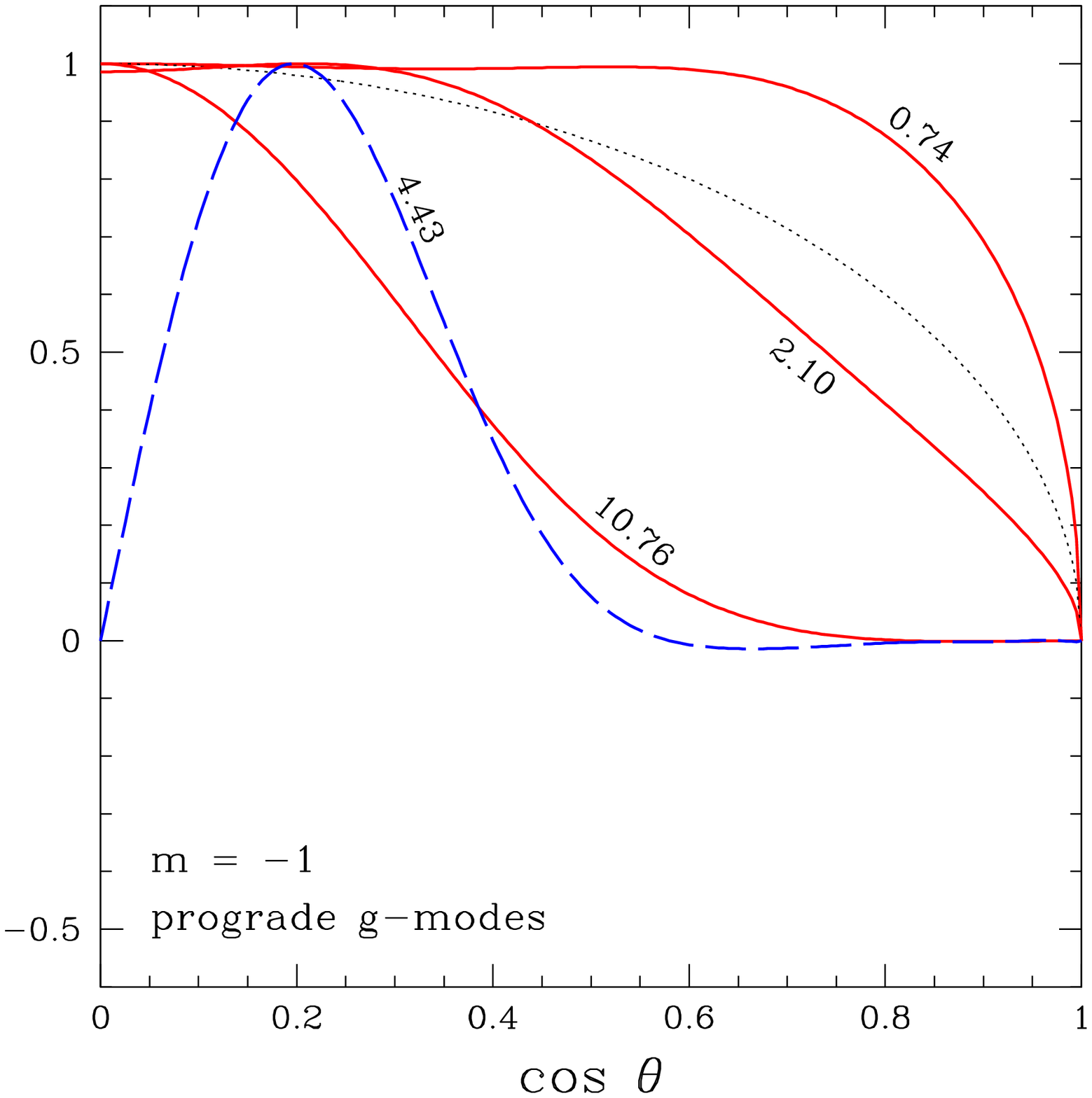}
\includegraphics[scale=.3]{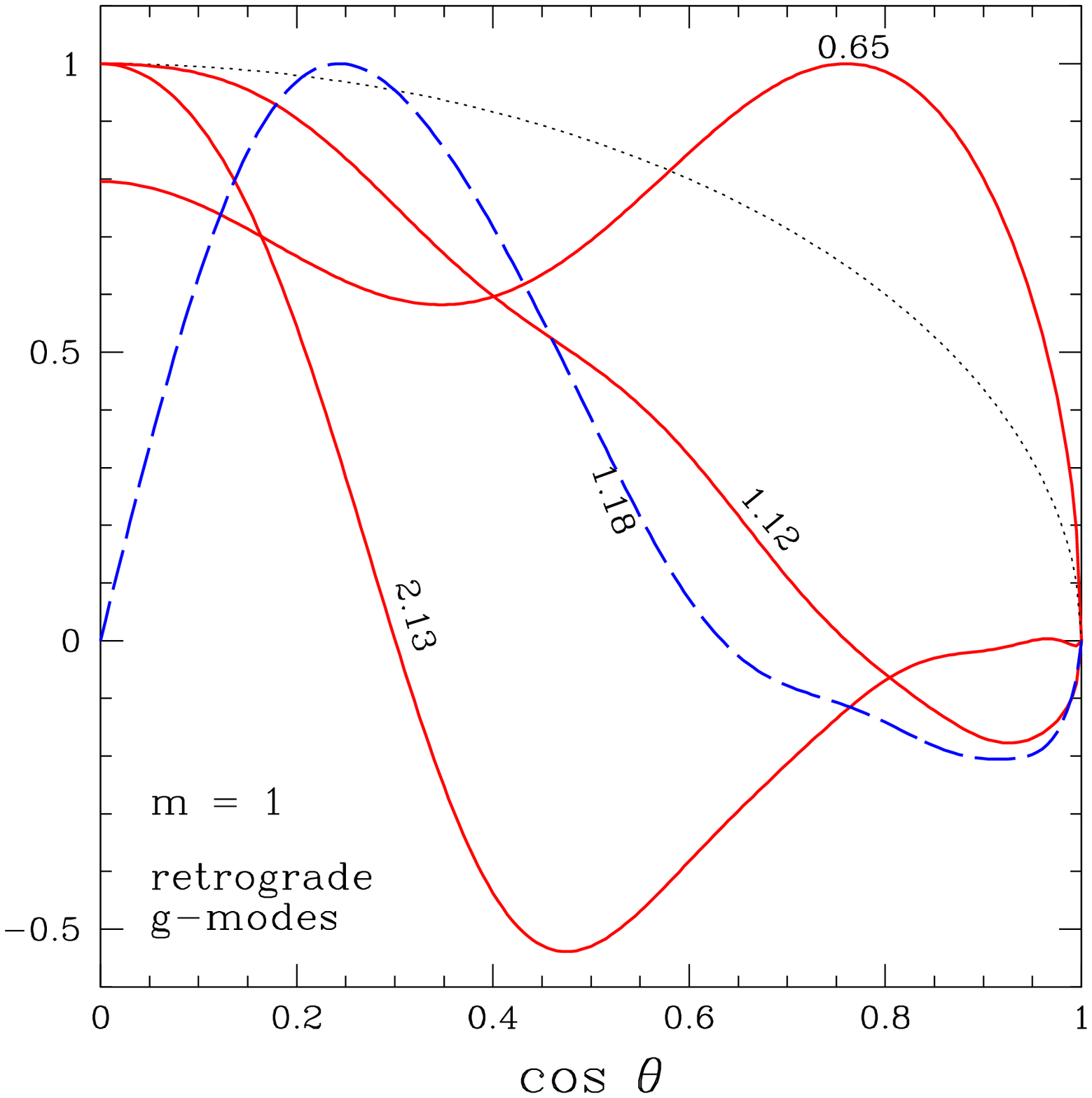}
\caption{
Amplitude distributions for scalar variables of g-modes with $m = \pm1$ and 
$\ell_0 = 1$ and $2$ as a function of $\cos\theta$, where  $\theta$ is co-latitude. 
Left and right panels are for prograde ($m=-1$) and 
retrograde ($m=1$) g-modes, respectively.
Solid and dashed lines are for even ($\ell_0=1$) and odd ($\ell_0=2$)
modes, respectively.
A number along each line indicates the value of $f_{\rm cor}=2\Omega/\omega$
for the mode. 
Note that
an additional nodal line appears for retrograde modes when $f_{\rm cor} > 1$,
which would reduce the visibility.  
Dotted line shows the associated Legendre function $P_1^1(\cos\theta)$. 
}
\label{fig:ampdist_gmodes}    
\end{figure}

Figure\,\ref{fig:ampdist_gmodes}  shows latitudinal distributions of the 
amplitude of $p'$ as a function of $\cos\theta$ for selected g-modes of 
$m=\pm 1$ and $\ell_0=1,~2$, in a $4.5M_\odot$ ZAMS model 
of rigid rotation at an angular frequency of 
$0.220 \sqrt{GM/R^3}$ (or with a period of 0.96\,days, $V_{\rm eq} = 130$km/s).
Although the adopted rotation rate is moderate,
the Coriolis force significantly affects 
the amplitude distribution of g-modes, because frequencies are so low that  
the parameter $f_{\rm cor}=2\Omega/\omega$ is larger than around $1$.
These eigenfunctions have been obtained by the method of Lee \& Baraffe~\cite{lee95} 
without using the `traditional' approximation 
in which the horizontal component of 
rotational angular velocity, $-\Omega\sin\theta \vec{e}_\theta$ is neglected. 
The results are very similar to those obtained by using the
traditional approximation (see, e.g., Lee \& Saio~\cite{lee97}), 
indicating that the approximation is well suited for 
low-frequency oscillations in which the horizontal velocity is 
much larger than the radial one.

Generally, amplitudes of oscillation modes with large $f_{\rm cor}$ tend 
to be confined to equatorial regions.  
This property is consistent with the behavior of the eigenvalue $\lambda$ of the 
Laplace tidal equation (e.g., Lee \& Saio~\cite{lee97} and Aerts et al.\ \cite{aer10}),
which governs angular dependence of oscillation in the traditional approximation.
As $f_{\rm cor}$ increases, $\lambda$ increases rapidly except for sectoral prograde
modes ($\ell_0 =-m$); a large $\lambda$ corresponds to a large effective degree $\ell$,
i.e., $\lambda \rightarrow \ell_0(\ell_0+1)$ as $\Omega \rightarrow 0$.

Solid lines in Fig.\,\ref{fig:ampdist_gmodes} are for $\ell_0=|m|=1$; i.e., 
sectoral modes, which have no latitudinal nodal lines when $\Omega=0$. 
The prograde modes ($m=-1$) keep this property 
even for a large $f_{\rm cor}$, while for retrograde modes ($m=1$) a latitudinal 
nodal line appears for the cases with $f_{\rm cor}>1$, which reduces visibility 
of the mode due to cancellation on the surface.

\begin{figure}[t!]
\hspace{-0.02\textwidth}
\includegraphics[scale=.3]{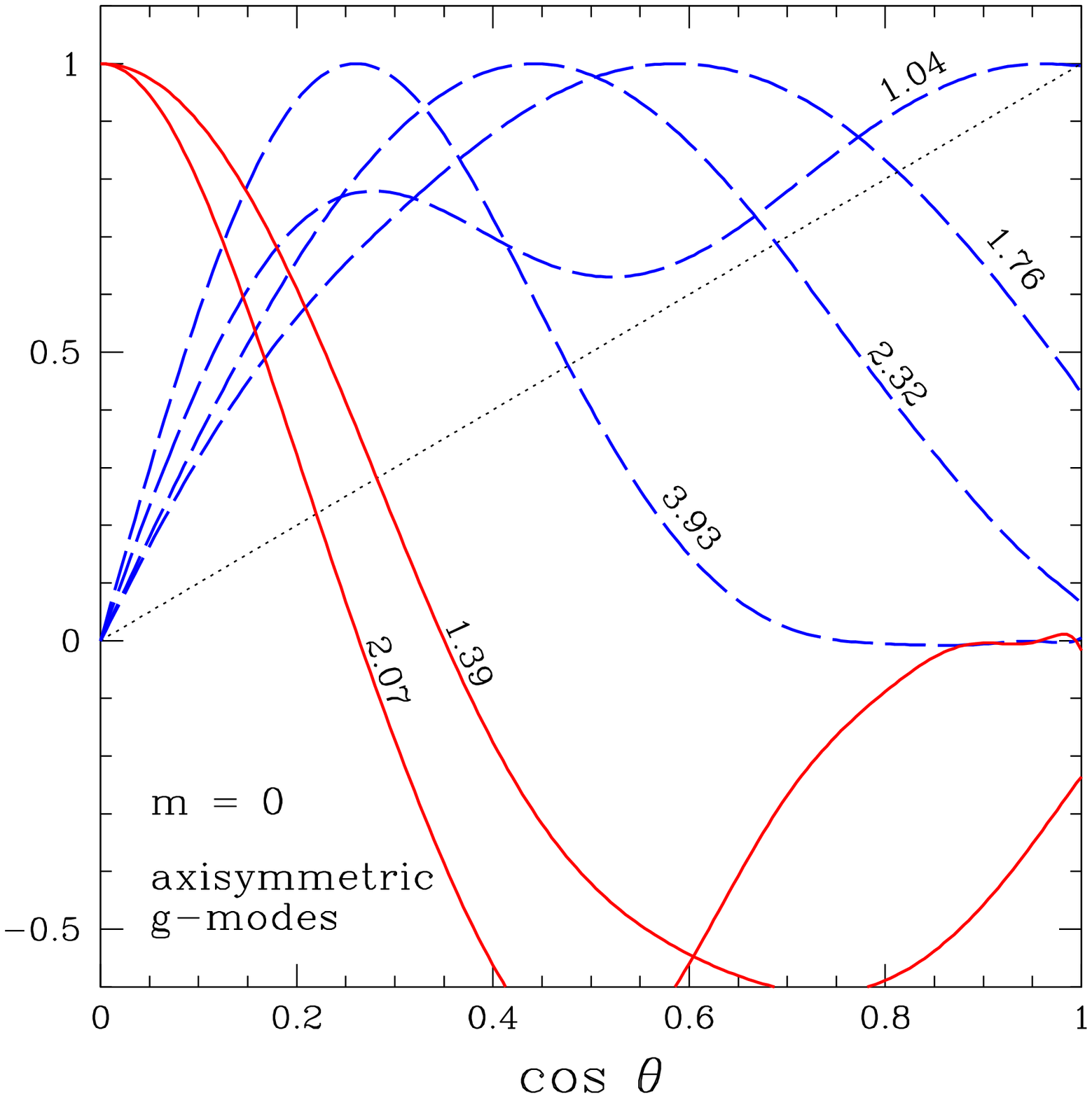}
\includegraphics[scale=.3]{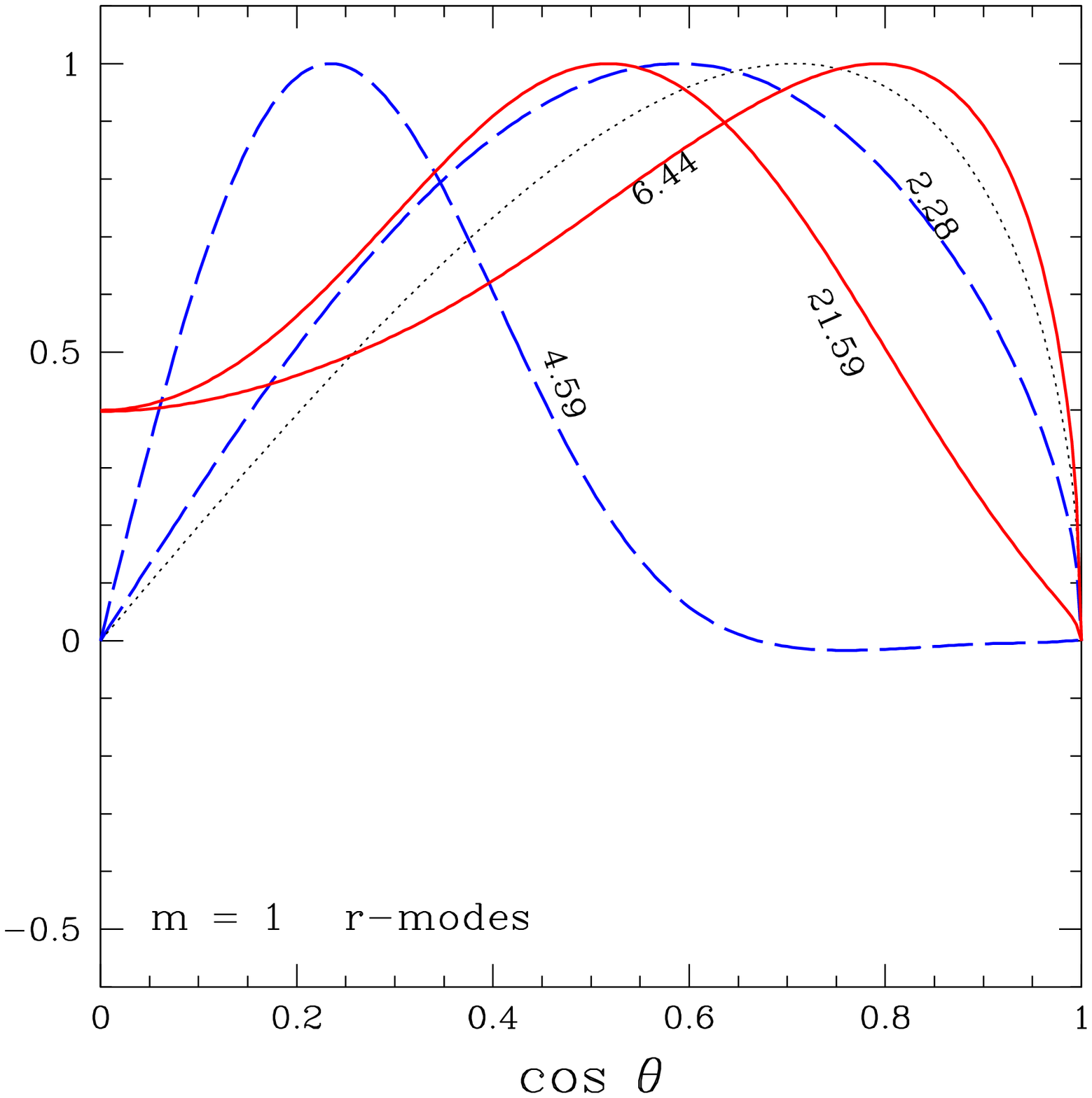}
%
\caption{
The same as Fig.\,\ref{fig:ampdist_gmodes} but for g-modes of $m=0$ (left panel)
and r-modes of $m = 1$ (right panel).
In the left panel, solid and dashed lines are $\ell_0=2$ (even) and $\ell_0=1$ 
(odd) modes, respectively.
In the right panel, solid lines are for even modes with $\ell=\ell'-1=1$,
and dashed lines are for odd modes with $\ell=\ell'+1=2$.
Dotted lines shows the Legendre function of $P_1(\cos\theta)$ (left panel)
and $P_2^1(\cos\theta)$ (right panel).
}
\label{fig:ampdist_rmodes}      
\end{figure}

The left panel of Fig.\,\ref{fig:ampdist_rmodes} shows latitudinal amplitude distributions
of axisymmetric ($m=0$) g-modes (dashed lines for $\ell_0 =1$ modes, 
and solid lines for $\ell_0=2$ modes). 
Similarly to non-axisymmetric modes, amplitudes of axisymmetric modes tend to 
be confined in equatorial zones as $f_{\rm cor}$ increases; in particular for  
$\ell_0=1$ modes the amplitude weight shifts from the pole to low-latitudes.

The right panel of Fig.\,\ref{fig:ampdist_rmodes} shows 
amplitude ($p'$) distributions of selected
r-modes of $m=1$ in the same model as in Fig.\,\ref{fig:ampdist_gmodes}.
Red solid lines are for even r-modes with $\ell'=\ell+1=2$ and 
blue dashed lines are for odd r-modes with $\ell'=\ell-1=1$, 
where $\ell'$ and $\ell$ are the latitudinal 
degrees for the dominant toroidal component and for the corresponding 
spheroidal component, respectively.
(The parity refers to the property of scalar variables.)

For r-modes of $m=1$ (right panel of Fig.\,\ref{fig:ampdist_rmodes}),  
a low-order mode close to the limiting frequency $\omega_{\rm rlim}$ 
(Eq.\,(\ref{eq:rlim})) 
and a relatively high-order mode are shown for each parity.
(For low-order r-modes of $m=1$, $f_{\rm cor} \approxsim \ell'(\ell'+1)$.)  
The amplitude distribution of a higher order r-mode tend to be more confined to
the equatorial region, which is similar to  high-order g-modes having large $f_{\rm cor}$.
An important difference from retrograde g-modes is that no additional latitudinal nodal-line
appears for r-modes.
The stability analyses for r-modes by Townsend~\cite{tow05} and Lee~\cite{lee06} 
indicate that odd r-modes are more easily excited in B-type main-sequence stars, 
while Savonije~\cite{sav05} found even (symmetric with respect to the equator) r-modes 
to be excited in some models.
We expect odd r-modes to be detected unless the inclination angle between
rotation axis and the line-of-sight is close to $90^\circ$. 

\subsection{Expected frequency ranges of g- and r-modes}
High order g-modes (Gautschy \& Saio~\cite{gau93}, Dziembowski et al.\ \cite{dzi93})
and r-modes (Townsend~\cite{tow05}, Savonije~\cite{sav05}, Lee~\cite{lee06}) 
are excited by the kappa-mechanism 
at the Fe-opacity bump in intermediate mass (around $3$--$8M_\odot$) main-sequence stars.
Some g-modes are, however, damped in rapidly rotating stars ($f_{\rm cor}> 1$) 
due to mode couplings.
The mode coupling occurs when two 
normal
modes with the same $m$ and parity but with 
different $\ell_0$ have similar oscillation frequencies, as these two modes 
are no longer independent in the presence of the Coriolis force.
Lee~\cite{lee01} has found that the mode couplings tend to damp retrograde 
g-modes.
The effects of the mode couplings are discussed in detail in 
Aprilia et al.\ \cite{apr11} and Lee~\cite{lee08,lee12}. 
The damping effects can also be significant on prograde g-modes except for 
sectoral modes ($\ell_0=-m$).
Although some prograde tesseral ($\ell_0 > -m$) g-modes might still remain  
excited in rapidly rotating stars, the effective degrees ($\ell$) of these modes 
might be too high to be detectable. 
Therefore, most visible modes should be low-degree prograde sectoral ($\ell_0=-m$) 
g-modes and (retrograde) r-modes in rapidly rotating intermediate mass stars. 

For high order g-modes in rapidly oscillating stars $\omega \ll \left|m\right|\Omega$ with 
$\omega$ being frequencies in the co-rotating frame so that we expect
to observe these modes at 
$\sigma = \left|\omega - m\Omega\right| \approxsim \left|m\right|\Omega$.

Since sectoral prograde g-modes are expected most visible as discussed above,
we expect groups of frequencies slightly above $\Omega$ and $2\Omega$ for $m=-1$ and
$-2$ modes, respectively. (Modes with higher $\left|m\right|$ are expected to be 
less visible due to cancellation.)

For r-modes with $m=1$ only odd modes ($\ell'=1$) seem to be excited in B-type stars, 
they have $\omega \approxsim \Omega$, whose observational frequencies are 
very small $\sigma \ll \Omega$. 
Symmetric (even) r-modes of $m=2$ ($\ell=2$) are also excited in some
cases, for which $\ell' =3$, hence $\omega_{\rm rlim}= {1\over3}\Omega$.
Therefore, frequencies in the observer's frame are slightly larger than ${5\over3}\Omega$.

In summary, we expect to observe groups of frequencies at approximately
\begin{equation}
\mbox{g-modes}: \quad \Omega, \quad 2\Omega, \quad \mbox{and r-modes}:
\quad 0,  \quad {5\over3}\Omega, 
\label{obsfreqs}
\end{equation}
where we have assumed arbitrarily that modes with effective degrees 
$\ell \le 2$ are visible.  

\begin{figure}[t!]
\begin{center}
\includegraphics[scale=.45]{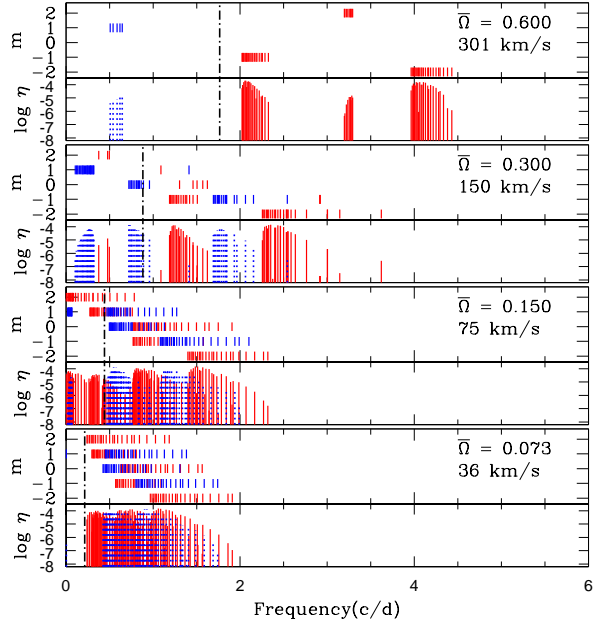}
\end{center}
\caption{Growth rates and azimuthal order $m$ of low-degree ($\ell \le 2$) 
modes excited in $4.5M_\odot$ main-sequence models with various rotation frequencies, where
the results were obtained by the method of Lee \& Baraffe~\cite{lee95}.
The cyclic rotation frequency is indicated by a vertical dash-dotted line at each panel, and 
$\overline{\Omega}$ is the corresponding angular frequency of rotation normalized 
by $\sqrt{GM/R^3}$ with $R$ being the mean stellar radius; equatorial rotation velocity is also shown. 
Horizontal axis indicates frequency in the observer's frame. 
}
\label{fig:vari_rot}
\end{figure}

Figure\,\ref{fig:vari_rot} shows growth rates and azimuthal order $m$ of the low degree
$(\ell \le 2)$ modes excited in a $4.5M_\odot$ main-sequence model of rigid 
rotation at various speeds.
The abscissa indicates frequencies in the 
observer's frame. Red (solid) and blue (dotted) lines are for even and odd modes, respectively.
In the model of the lowest rotation frequency 
($\overline{\Omega}=0.073$, bottom panel), no r-modes are excited, and we see no 
appreciable effects of rotation on the stability of g-modes; 
rotation only disperses frequencies by the effect of transformation from co-rotating 
to inertial frames; $\sigma = \omega - m\Omega$. 

In the second lowest rotation case with a normalized angular frequency of rotation
$\overline{\Omega}=0.15$, $m=1$ odd r-modes with $\ell'=\ell-1=1$ are excited;
they have very small frequencies in the observer's frame, because 
$\omega_{\rm rlim} = \Omega$ (Eq.\,(\ref{eq:rlim})).

As the rotation frequency increases ($\overline{\Omega} > 0.2$), retrograde 
($m>0$) g-modes tend to be damped, while prograde g-modes remain excited. 
Due to the damping of retrograde g-modes and the increasing effect of $-m\Omega$,
frequency groupings become conspicuous as the rotation frequency increases. 
In a sufficiently rapidly rotating case (as in the top panel), well populated 
frequency groups of prograde sectoral ($\ell=-m$) g-modes 
are formed around $1.2\Omega$ and $2.3\Omega$ corresponding to $m=-1$ and $-2$, 
respectively.
 
The r-modes also form groups, because the frequency deviation 
$\left|\omega_{\rm rlim}-\omega\right|$ of an r-mode
is usually much smaller than $m\Omega$.
The group  at smallest frequencies corresponds to $m=1$ anti-symmetric r-modes
$\ell'=\ell-1=1$. 
In the fastest rotation case (top panel) all the excited retrograde ($m>0$) modes  are r-modes. 
Note that in this model $m=2$ r-modes with $\ell=\ell'-1=2$ symmetric with respect
to the equator are excited; their observational frequencies are slightly larger than
$\left|\omega_{\rm rlim} - m\Omega \right| = {5\over3}\Omega$.
The excitation of symmetric $m=2$ r-modes in a sufficiently 
rapid rotator agree with the results of Savonije \cite{sav05}.

Figure\,\ref{fig:vari_rot} shows that frequency groupings appear even in stars whose
rotation rates are considerably less than the critical rate.
This is consistent with 
the finding of Balona et al.\ \cite{keplerB} from Kepler data that
several B stars with relatively large $V\sin i$ show frequency groupings.

\section{Comparisons with frequency groups of  Be stars}

The characteristic frequency groupings obtained for several Be stars
(Fig.\,\ref{fig:befreq} ) were fitted with models 
(e.g., Cameron et al.\cite{cam08}, Neiner et al.\cite{nei12}).
Such fittings give estimates for 
rotation frequencies and
in some cases, the constraint to the internal mixing in rapidly rotating stars.
In this section, we discuss what we learn from model fittings to low-frequency oscillations
of Be stars.

\subsection{HD\,50209}
\begin{figure}[t!]
\hspace{-0.04\textwidth}
\includegraphics[scale=.35]{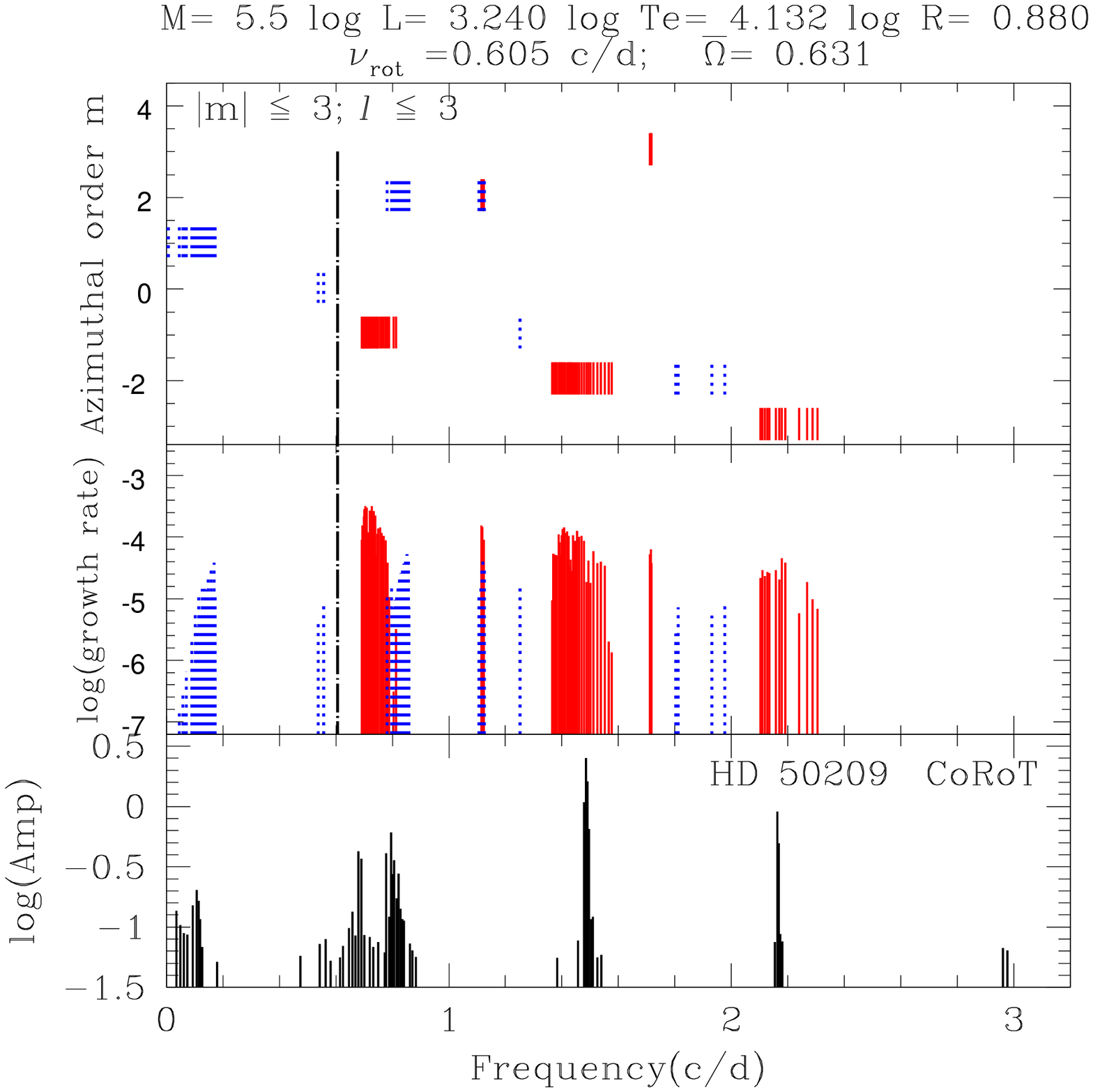}
\includegraphics[scale=.26]{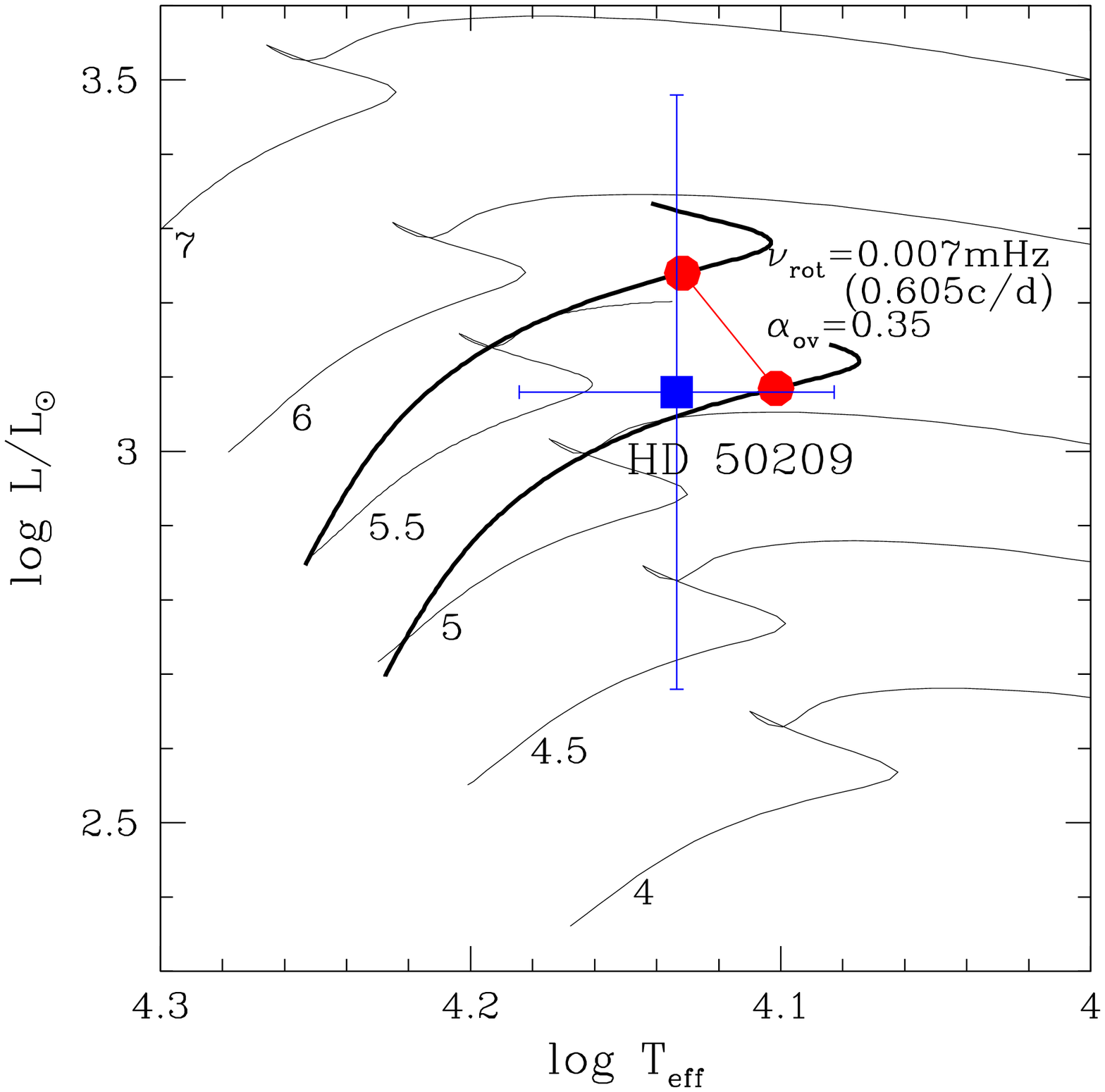}
\caption{
{\bf Left panel:}
Frequencies of excited modes in a model of a $5.5M_\odot$ star rotating at a rate 
of $0.605$\,c/d are compared with observed frequencies of HD\,50209 (bottom panel) 
adopted from Diago et al.\ \cite{dia09}; 
the rotation frequency of the model is indicated by the dash-dotted vertical line.
The top panel shows azimuthal order $m$ (prograde modes correspond to $m<0$) and
the middle panel shows growth rates of excited modes. 
Solid (red) and dashed (blue) lines are even and odd modes respectively. 
{\bf Right panel:}
The positions of HD\,50209 (filled square with error bars) on the HRD and two models 
(filled circles) that are reasonably consistent with the observed frequency groups; 
a fit by one of the models is shown in the left panel.
Evolutionary tracks are calculated with a standard composition of $(X,Z)=(0.70,0.02)$.
The thick two tracks for $5.0M_\odot$ and $5.5M_\odot$ are from evolution
including overshooting of $0.35H_p$ around the convective core; the other
tracks are without overshooting.  
}
\label{fig:hd50209}
\end{figure}

We will discuss, as an example, a model fitting to low-frequency oscillations of the 
late-type (B8IVe) Be star HD\,50209 detected by CoRoT (\cite{dia09}); 
the amplitude spectrum consists of five 
frequency groups around $0.1$c/d, $0.6$--$0.8$c/d, $1.5$c/d, $2.2$c/d and $3$c/d
(or six groups if the $0.6$--$0.8$c/d group is separated into two groups)
as shown in Fig.\,\ref{fig:befreq} and in the left-bottom panel of Fig.\,\ref{fig:hd50209}. 

The fundamental parameters of HD\,50209 have been derived by Diago et al.~\cite{dia09} as
$\log T_{\rm eff} =  4.134\pm 0.051$, $\log L = 3.02\pm0.39$, and $\log g = 3.56\pm0.11$
(with $\Omega/\Omega_{\rm crit} = 0.90$).  
The estimated position of HD\,50209 in the HRD is shown in the right panel of 
Fig.\,\ref{fig:hd50209} by a filled square with error bars.
At the center of the parameter ranges the critical rotation rate is around $0.7$c/d.
Since prograde g-modes of $m=-1$ form  a group around frequencies slightly
larger than the rotation frequency, 
the $0.6$--$0.8$c/d group should be fitted by assuming a rotation frequency of 
approximately $0.6$c/d. 
The rotation frequency should not be far from the critical rate; i.e., the radius 
should be sufficiently large, otherwise no 
clear frequency groupings would be expected (see Fig.\,\ref{fig:vari_rot}).
Two models of $5.0M_\odot$ and $5.5M_\odot$ stars rotating at a rate of 
$0.6$\,c/d ($0.007$\,mHz)  were examined;  their positions
on the HRD are shown by filled circles in Fig.\,\ref{fig:hd50209}.
In both cases a substantial core overshooting (more than $0.3H_p$) must be 
included to meet the above requirements for 
$T_{\rm eff}$ and luminosity within the spectroscopic estimates.
The requirement of substantial core overshooting for the model of HD\,50209 
is common to the models for other CoRoT Be stars, HD\,175869 and HD\,181231
as discussed in Neiner, et al.\ \cite{nei12}.

Both models produce similarly good fits to observed frequency groups of HD\,50209.
A comparison with the $5.5M_\odot$ model is shown in the left panel of 
Fig.\,\ref{fig:hd50209}, where the method of Lee \& Baraffe~\cite{lee95} was used
for the stability analysis.
To fit the three ($0.6$--$0.8$, around $1.5$ and $2.2$\,c/d) groups with g-modes,
we have to consider azimuthal orders up to 3 ($\left|m\right|\le 3$). 
(The frequencies at around $3$\,c/d are probably harmonics of large amplitude frequencies
at approximately $1.5$\,c/d as indicated by Diago et al~\cite{dia09}.
) 

One can interpret the smallest frequency group at around $0.1$\,c/d as due to $m=1$ 
anti-symmetric r-modes.
Diago et al.~\cite{dia09} estimate an inclination angle of $i\sim 60^\circ$ for HD\,50209 for
a rotation frequency of $90$--$95$\% of the critical rate.
With this inclination angle, r-modes would suffer from no serious surface cancellation 
(see right panel of Fig.\,\ref{fig:ampdist_rmodes}),
and would be detectable if the r-modes produce enough temperature variations.  
It is interesting to note that a very low-frequency group is 
common in SPBe stars that show g-mode pulsations, but it does not appear in 
hot Be stars ($\zeta$ Oph and HD\,49330) with p-mode pulsations (Fig.\,\ref{fig:befreq}).
This agrees with the theoretical prediction for  the excitation range of r-modes on the HRD 
as obtained by Townsend~\cite{tow05}, and supports the identification of the very low
frequency groups in SPBe stars as r-modes.
In this model, symmetric r-modes of $m=2$ and 3 are excited, but they have no 
observational correspondence.

\subsection{Rotation rates of Be stars}

\begin{figure}[t!]
\begin{center}
\includegraphics[scale=.45]{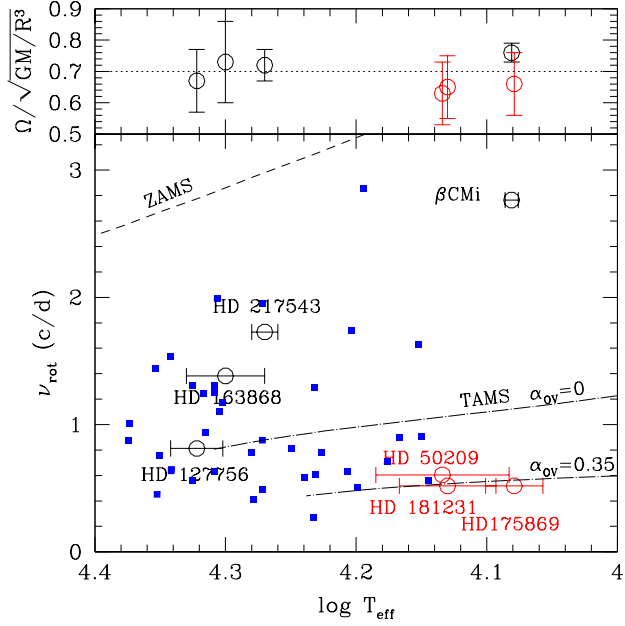}
\end{center}
\caption{
Effective temperature versus rotation frequencies estimated by model fittings for
Be stars observed by MOST and CoRoT (large circles with error bars).
The top panel shows rotation frequency normalized by $\sqrt{GM/R^3}$, where
$R$ is the mean radius. With this normalization the
critical rotation frequency corresponds to approximately $0.7$. 
Filled squares are $\lambda$ Eri variables whose rotation frequencies are assumed to be
$80$\% of the observed frequencies listed in Balona \cite{bal95}.
The dashed line indicates the critical rotation for zero-age main-sequence models 
as a function of $\log T_{\rm eff}$.
Dotted-dashed lines show critical rotation of TAMS (terminal-age main sequence) models
without and with substantial convective core overshooting. 
}
\label{fig:terot}
\end{figure}

Figure\,\ref{fig:terot} plots rotation frequencies obtained by fitting theoretical 
predictions for frequency groups against those observed 
for several Be stars observed by MOST and CoRoT
(a similar diagram is shown in \cite{cam08}).
The lower panel shows the rotation frequency of the Be stars (with error bars) as a function
of the effective temperature.
The Be stars observed by CoRoT tend to have low rotation frequencies compared to those 
observed by MOST.
This is due to the fact that CoRoT-observed Be stars tend to be more evolved, having 
spectral luminosity class IV. 
This apparent difference disappears if the rotation frequency is normalized by 
$\sqrt{GM/R^3}$ as plotted in the top panel of Fig.\,\ref{fig:terot}, where
$R$ is mean radius.
With this normalization critical rotation corresponds to  $\Omega/\sqrt{GM/R^3}\approxsim 0.7$.
This indicates that all Be stars shown can be considered rotating nearly critically, 
which agrees with recent spectroscopic estimates by Townsend et al.\ \cite{tow04} and
Cranmer~\cite{cra05}. 

Dashed and dot-dashed lines in the bottom panel of Fig.\,\ref{fig:terot}  show critical 
rotation frequency as a function $\log T_{\rm eff}$ for ZAMS and TAMS (terminal age 
main-sequence) models, respectively.
For TAMS models, critical rotations with and without substantial core overshooting cases 
are shown. 
Obviously, substantial overshooting would need to be assumed for models to
be consistent with the CoRoT observations of Be stars.

Also plotted in this figure are rotation frequencies of $\lambda$ Eri variables (filled squares);
a group of Be stars which show short-term (order of a day) periodic light variations known
from ground-based observations (e.g., Balona~\cite{bal95}). 
Folded light curves of $\lambda$ Eri variables, some of which show double-wave character,
are consistent with the grouped frequency distributions obtained for 
relatively late-type Be stars monitored by MOST and CoRoT satellites.
In fact, the folded light curve of HD\,181231 from CoRoT observations 
(Neiner et al.\ \cite{nei09}), for example,
is similar to typical light curves of $\lambda$ Eri variables. 
Since a model with a rotation frequency about 20\% smaller than the frequency 
at the group of $m=-1$ modes
fits well to such a characteristic frequency
distribution, rotation frequencies of $\lambda$ Eri variables are assumed as 
80\% of the observed frequencies, and those values are plotted in Fig.\,\ref{fig:terot}. 
This figure shows that the distribution of $\lambda$ Eri rotation rates are consistent with
Be star models rotating nearly critically if a substantial core-overshooting of around 
$0.35H_p$ is assumed to occur in these stars.
Since models without core overshooting are consistent with the slowly rotating SPB stars,
rapid rotation seems the cause of the substantial 
mixing needed around the convective core (Neiner et al.\ \cite{nei12}).

\section{Conclusion}
Rapid stellar rotation modifies properties of oscillations in complex ways and 
makes comparisons between observational and theoretical results difficult.
In particular, low order p- and g-modes excited in early type Be stars in the $\beta$ Cephei
instability region show complex frequency spectra affected considerably both by 
deformation of the equilibrium structure and by the Coriolis force. 
The frequency distributions of these stars are such that it would
seem impossible to identify modes and so compare with theoretical results.

In contrast to the p-modes, rapid rotation helps us to 
identify the azimuthal degree $m$ of 
low-frequency modes, because observational frequencies of these modes form groups;
i.e.,  g-modes with the same $m$ have similar frequencies. 
This property can be used to identify $m$ and to estimate the rotation frequency of each star.
The rotation rates thus obtained for several
Be stars indicate that Be stars rotate nearly critically, which confirms conclusions drawn
from spectroscopic analyses. 
In addition, in order to fit observations of some Be stars, models with substantial 
core-overshooting are needed. 
This indicates that extensive internal mixing exterior to the convective core
occurs in rapidly rotating stars as discussed in Neiner et al.\ \cite{nei12}.
Furthermore, it might be possible in the near future to compare each g-mode (and r-mode) 
frequency (or period spacings) with theory in order to obtain detailed information
on the stellar interior. 
Rapid rotations are advantageous in this case because we can identify the 
azimuthal order $m$ while effects of centrifugal deformation remain small, as shown by 
Ballot et al.\ \cite{bal11}.

\begin{acknowledgement}
I am grateful to Alfred Gautschy and Umin Lee for helpful comments.
\end{acknowledgement}

\end{document}